\documentclass[12pt,preprint]{aastex}
\usepackage{amsmath,amssymb,amsfonts,longtable,rotating}
\usepackage{epsfig}                   %%
\usepackage{epstopdf}                   %%
\usepackage{natbib}
\usepackage{txfonts}                  %%
\usepackage{graphicx}
\usepackage{lscape}
\usepackage{rotating}

\shorttitle{Chemical analysis of CD$-$27 14351}
\shortauthors{Karinkuzhi et al.}
\begin{document}

\title{Chemical analysis of a carbon-enhanced very metal-poor star: CD$-$27 14351}

\author{Drisya Karinkuzhi\altaffilmark{1}
\affil{1 - Indian Institute of Astrophysics, Koramangala,
Bangalore 560034}}

\and

\author{Aruna Goswami\altaffilmark{1},
\affil{1 - Indian Institute of Astrophysics, Koramangala,
Bangalore 560034}}

\and

\author{Thomas Masseron\altaffilmark{2},
\affil{2 -  Institute of Astronomy, Madingley Road, Cambridge CB3 0HA, UK}}

\begin{abstract}

We present the first time abundance analysis of a very metal-poor
 carbon-enhanced  star CD$-$27 14351 based on a high resolution
(R ${\sim}$ 48\,000) FEROS spectrum. Our abundance analysis performed
using Local Thermodynamic
Equilibrium (LTE) model atmospheres shows that the object is a cool star with
stellar atmospheric parameters, effective temperature T$_{eff}$ = 4335 K, 
surface gravity log\,g = 0.5, microturbulence 
$\xi$ = 2.42 km s$^{-1}$, and, 
metallicity [Fe/H] = $-$2.6. The star exhibits high carbon and
nitrogen abundances with [C/Fe] = 2.89 and [N/Fe] = 1.89. Overabundances of
neutron-capture elements are evident in Ba, La, Ce, and  Nd  with estimated
[X/Fe] $>$ 1, the largest enhancement being seen in Ce with [Ce/Fe] = 2.63.
While the first peak s-process
elements Sr and Y are found to be enhanced with respect to Fe, ([Sr/Fe] = 1.73
and [Y/Fe] = 1.91) the third peak s-process element Pb could not be detected
in our spectrum at the given resolution.
 Europium, primarily a r-process element also  shows an  enhancement with 
[Eu/Fe] = 1.65.  With [Ba/Eu] = 0.12 the object CD$-$27 14351  satisfies the
classification criterion for  CEMP-r/s star.
The elemental abundance distributions observed in this  star is  discussed
in light of  
chemical abundances observed in other CEMP stars from literature.

\end{abstract}

\keywords{  stars: carbon ----
 stars: chemically peculiar---- stars: late-type ---- stars: low-mass }

\section{INTRODUCTION}
The object CD$-$27 14351 drew our attention when a low-resolution spectrum of
this object acquired on Oct 23, 2015, during the observing run of our
observational program with HCT (Himalayan Chandra Telescope, at IAO, Hanle;
 one of the objectives of this program  is  to search for CH stars and
carbon-enhanced metal-poor stars among field stars)  on visual
inspection was found to show significant resemblances with the low-resolution
spectrum of  HD~5223, a well-known classical CH star. From the known stellar
parameters of HD~5223,
(T$_{eff}$ = 4500 K, log\,g = 1.0 and metallicity [Fe/H] = $-$2.06
\citep{goswami06}, a first guess  was  the object
CD$-$27~14351  could also  be a  cool metal-poor object similar to HD~5223.
Dominant carbon bearing molecular bands in its spectrum  also indicated that
the object is likely carbon-enhanced. 
For an  understanding of its  elemental abundance pattern and its origin,
 a detailed chemical composition study was taken up based on
a high resolution FEROS (Fiber-fed Extended Range Optical Spectrograph,
connected to the 1.52 m telescope at Chile) spectrum.

Our analysis reveal  that the object is indeed a carbon-enhanced metal-poor
star that also show significant enhancement of  nitrogen and neutron-capture
elements. The  heavy elements abundance estimates  indicate that the
surface composition of the object has significant contributions  originated
from both the r-process as well as the s-process nucleosynthesis.
 This is supported  by our estimate of the  abundance ratio
 [Ba/Eu] (=0.12),  that   qualifies the object to be placed in the
CEMP-r/s
group (\citet{beers05}, for definition of CEMP-r/s stars and other
CEMP stars). However,
the origin of CEMP-r/s stars still remains far from clearly understood.
Several proposed physical scenarios are available  in  literature \citep{qian03,zijlstra04,barbuy05,wanajo06}.  In
spite of considerable efforts no single  mechanism has yet been identified
that could explain satisfactorily the production of enhanced carbon and
heavy-element abundance patterns observed  in the group of these stars. While 
the enhancement in the s-process elements  is believed
to be due to mass transfer in binary systems involving  AGB companion
that underwent s-process nucleosynthesis, the origin of the  enhanced
abundance of the r-process element Eu  remains unexplained in this scenario.
In a recent work \citet{hampel16} have shown that the observed abundance
patterns of CEMP-r/s stars could  be convincingly reproduced through i-process 
(intermediate neutron-capture process) which  operates at neutron densities
between those of the s-process and r-process. Except for one object 
the abundance
patterns of all the stars in their sample could be reproduced by this process.
To place observational constraints on the nucleosynthesis of s-,  r- and 
i-process
elements at low metallicity, it is necessary to conduct high-resolution
spectroscopic analysis  of stars with excess of heavy elements and
 CEMP stars undoubtedly form  important targets.
Such studies  would help provide insight into the possible origin of these
classes  of objects.
Our detailed  high resolution spectroscopic analysis of CD$-$27 14351 is
prompted by such  motivations.

In Section 2, we describe the source  of the low-resolution
and high resolution  spectra of the
programme star.  In  section 3  we present our estimates of effective
temperatures based on  available BVJHK photometry of the object.  The
spectral analysis procedures and estimation of the stellar atmospheric
parameters are discussed in Section 4. The elemental abundance results
and discussions are presented in Section 5. Uncertainty in elemental
abundances are discussed in section 6. 
 Conclusions are provided in Section 7.

\section{ Spectra of CD$-$27 14351 }
The low-resolution  spectrum of CD$-$27 14351 was acquired
on October 23, 2015, using HFOSC attached to the  2-m HCT (Himalayan
Chandra Telescope)  at the Indian Astronomical Observatory
(IAO), Mt Saraswati, Digpa-ratsa Ri, Hanle. The spectrograph used is HFOSC
(Himalayan Faint Object Spectrograph Camera). The spectrum covers the
wavelength range 3800 - 6800 \AA\, at a resolution R ${\sim}$ 1300.
The high resolution  FEROS (FEROS: the Fiber-fed Extended Range Optical
Spectrograph, connected to the 1.52 m telescope at ESO, Chile) spectrum
of CD$-$27 14351 used in this study was acquired on July 14, 2000, 
has a resolution of   R ${\sim}$ 48000
and covers the wavelength  range spanning  3500 - 9000  \AA\,. The estimated
radial velocity  is ${\sim}$ 61.1 km s$^{-1}$.
The basic data for this  object is  listed in  table 1.

\begin{figure}
\centering
\includegraphics[height=12cm,width=12cm]{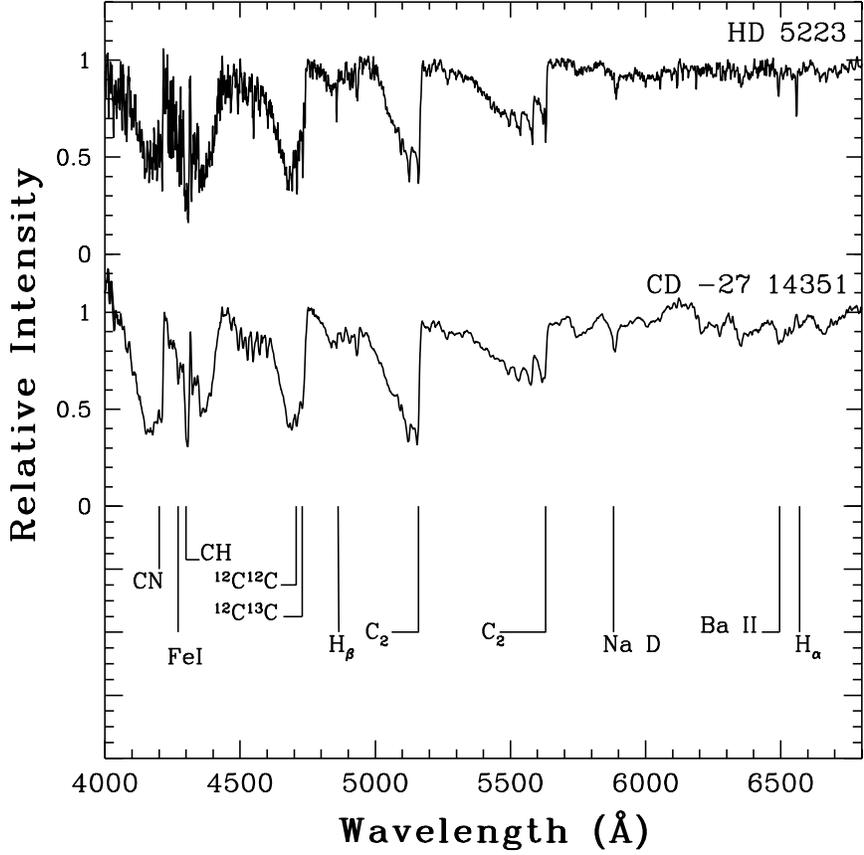}
\caption{A comparison of the low resolution spectrum of CD$-$27 14351 with the 
spectrum of   CH star HD~5223. Some prominent features are indicated in the 
figure. }

%\label fig1
\end{figure}

\begin{table*}
\centering
{\bf Table 1: Basic data for the program stars}\\
\begin{tabular}{cccccccc}
\hline
Star Name.   & RA(2000) & DEC(2000)& B& V&J&H&K \\
\hline
 CD$-$27~14351 &19 53 08.00 & -27 28 14.96&    11.82  & 9.70  &  7.022 &  6.296 &  6.135\\ 
\hline
\end{tabular}
\end{table*}

{\footnotesize
\begin{table*}
\centering
{\bf Table 2: Temperatures from  photometry }\\
\begin{tabular}{cccccc}
\hline
Star Name     &  $T_{eff}$ (J - K)  &  $T_{eff}$(J - H)  &  $T_{eff}$ (V - K) & $T_{eff}$ (B - V)& Spectroscopic  \\
              &     K             &        K            &     K              &      K    & K       \\
\hline
\\
  CD$-$27~14351&  4067 &  4026 (-2.0) &   3900 (-2.0) &   2635(-2.0)  & 4335\\
            &          &  3989(-2.5)&    3900 (-2.5) &   2528(-2.5)\\
           &          &  3938(-3.0)&    3904 (-3.0) &   2438(-3.0)\\
\hline
\\
\end{tabular}

The numbers in the parenthesis indicate the metallicity values at which the temperatures are calculated.\\
\end{table*}
}
 \section{ Temperatures from photometric data}
We have calculated the photometric temperatures of CD$-$27~14351 using the 
temperature calibrations of \citet{alonso99,alonso01}  for the giants.  
The estimated 
values  along with the spectroscopic temperature estimate are given in 
table 2.  As can be seen in the following section the derived 
spectroscopic temperatures are   
higher by $\approx$ 300 K from J - K and J - H temperatures and  
$\approx$ 400 K from the  V - K temperatures. This discrepancy is probably
due to  the high carbon  enhancement that changes significantly the colours; 
  due to the severe blending 
from the molecular lines B - V temperatures 
could be  much lower than the spectroscopic temperature estimate.

\section{Spectral analysis}
In figure 1, we show a comparison of  the spectrum  of CD $-$ 27~14351 with 
the spectrum of the  classical CH star HD 5223
\citep{Goswami05} obtained from  the same observational set up.
The spectra of these two objects distinctly  show close
resemblances with each other.
 The high resolution FEROS spectrum is used for determination of stellar 
atmospheric parameters and elemental abundances. The stellar parameters 
are determined by measuring    
 the equivalent widths of clean unblended Fe I and Fe II lines. The spectrum of 
CD$-$27~14351 is severely blended with contributions from  carbon molecular bands throughout 
the spectrum making it  difficult to find  clean unblended lines of  
Fe I and Fe II. Only seventeen lines of Fe I and two lines of  Fe II are found
usable  for deriving the stellar atmospheric parameters.  Lines with excitation 
potential in the range  0.0 eV to 5.0 eV and equivalent 
width 20 \AA\ to 160 \AA\  are selected for this purpose. 
These  lines  along with the abundances derived 
from each  line computed using the latest version of MOOG \citep{sneden73} are 
listed in table 3. We have used the Kurucz grid of model atmospheres with no 
convective overshooting  (http: //cfaku5.cfa.harvard.edu/) and assumed 
LTE condition  throughout the analysis.  
The method of excitation equilibrium is used for deriving the effective 
temperature $T_{eff}$. The initial value is assumed to be  the photometric 
temperatures and the  finally adopted value is obtained by an iterative 
process until the slope of the 
abundance versus excitation potential of Fe I lines is found to be nearly zero 
(Figure 2, lower panel).  
The microturbulent velocity is fixed at this temperature by demanding 
that there be no dependence of the derived Fe I abundance 
on the equivalent widths of the corresponding lines (Figure 2, upper panel).
 The surface gravity is fixed at a value which makes 
the abundances of iron from the Fe I and Fe II lines equal. 
The adopted stellar parameters are listed in table 4.       

\begin{figure}
\centering
\includegraphics[height=12cm,width=12cm]{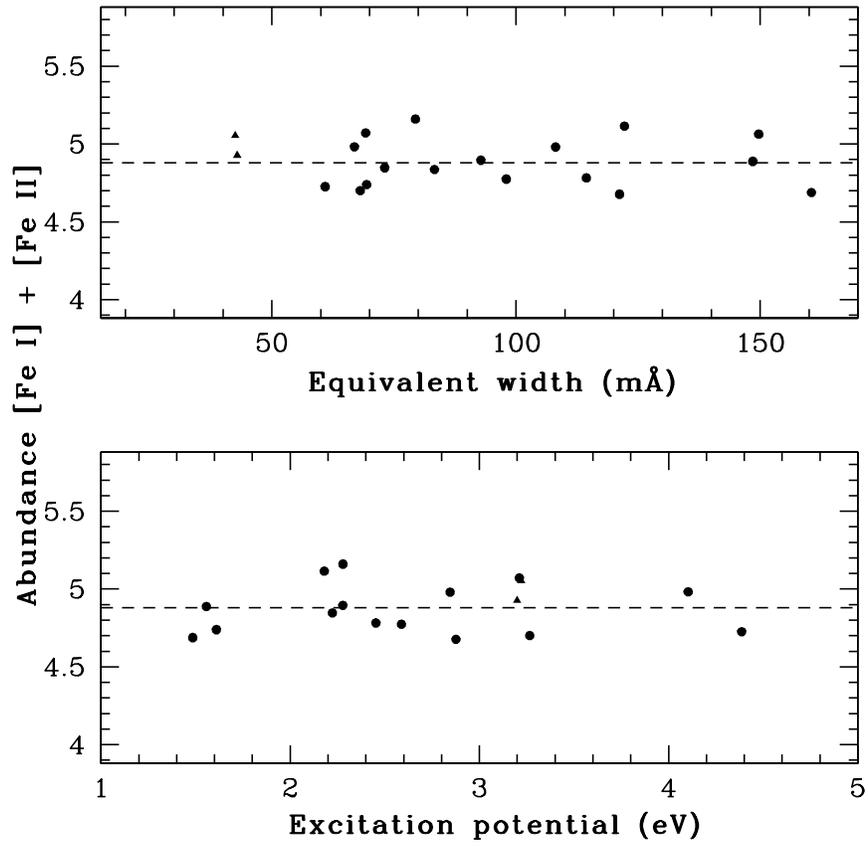}
\caption{The iron abundance of CD$-$27~14351 is shown for individual 
Fe I (solid circles) and Fe II (solid triangles) lines as a function of 
excitation potential (lower panel) and as a function of 
equivalent width (upper panel).}
%label fig2
\end{figure}

\begin{table*}
\centering
{\bf  Table 3. Lines  used for the analysis}\\
\begin{tabular}{lcccccc}
\hline
Wavelength&    Element    &    E$_{low}$ &   log gf&  Equivalent & log $\epsilon$& Reference\\
 ~~~~\AA\        &                &     ev           &            &      width (m\AA\ ) &dex &\\
\hline
  5688.220 &  Na I &  2.104 & -0.45  &  58.3  &     4.89& 1  \\
  4571.100 &  Mg I &  0.000 & -5.69  & 154.1  &     5.88&2 \\ 
  5528.405 &       &  4.350 & -0.62  &  61.1  &     5.04& 3 \\ 
  5172.680 &       &  2.710 & -0.40 &  304.7  &     5.92&4 \\ 
  5261.710 &  Ca I     & 2.521&  -0.73 &   27.1 &   4.39&5\\
  6439.070 &          &2.530 &  0.47  & 109.6 &    4.42 &5\\
  6499.650 &          & 2.523  &-0.59   & 65.2   &  4.72& 5 \\ 
  4999.500 & Ti I   & 0.830 &  0.25   &134.0  & 3.33&  6\\
  5210.390 &    &  0.048 & -0.88 &  120.6  &     3.04&6 \\ 
  5192.970 &    &0.020 &  -1.01 &   143.1& 3.38&6\\ 
  4443.790 &  Ti II   &1.080  &-0.70  & 178.7    & 3.29&6  \\  
  4589.950 &     & 1.236  &-1.79 &   94.9     &  3.02 &6\\
  5226.530 &       &  1.570 & -1.30 &  131.9  &     3.30&6\\ 
  5247.570 &  Cr I &  0.961 & -1.64  &  29.0  &     2.95&6 \\ 
  5348.312 &       &  1.003 & -1.29  &  33.0  &     2.72& 6\\ 
  4476.019 &  Fe I & 2.845 & -0.57  & 108.1  &     4.98&7 \\ 
  4890.755 &   &  2.876 & -0.43 &  121.2  &     4.68&8 \\ 
  4924.770 &   &  2.279 & -2.22  &  79.4  &     5.16& 8\\ 
  4982.524 &   &  4.103 &  0.14  &  66.9  &     4.99& 9\\ 
  4994.130 &   &  0.915 & -3.08  & 149.7  &     5.06&8 \\ 
  5171.600 &   &  1.485 & -1.79 &  160.5  &     4.69& 8 \\ 
  5194.941 &   &  1.557 & -2.09 &  148.5  &     4.89&8 \\ 
  5198.711&    &  2.223 & -2.14 &   73.1  &     4.85&8 \\ 
  5202.340 &   &  2.180 & -1.84 &  122.2  &     5.12&8\\ 
  5217.389 &   &  3.211 & -1.10&    69.2  &     5.07& 9\\ 
  5247.050 &   &  0.087 & -4.98&    83.3  &     4.84&8\\ 
  5307.370 &   &  1.610 & -2.98&    69.4  &     4.74&8 \\ 
  5339.928 &   &  3.266 & -0.68 &   68.1  &     4.70&8 \\ 
  5415.192 &   &  4.386 &  0.50 &   60.9  &     4.73& 8\\ 
  6137.694&    &  2.588 & -1.40 &   98.0  &     4.77&8 \\ 
  6136.615 &   &  2.453 & -1.40 &  114.4  &     4.78&8  \\
  6421.350 &   &  2.278 & -2.03 &   92.8  &     4.90& 8 \\
  5234.625 &   Fe II   &   3.22 & -2.05 &   38.5  &   4.97& 8 \\
  5276.002 &        &   3.199 & -1.90 &   40.5  &     4.88&8 \\

   \hline
\end{tabular}
\end{table*}

{\footnotesize

\begin{table*}
\centering
{\bf  Table 3 (cont.):  Lines used for the analysis}\\
\begin{tabular}{lcccccc}
\hline
Wavelength&    Element    &    E$_{low}$ &   log gf& Equivalent  &log $\epsilon$&References\\
 ~~~~\AA\        &                &     ev           &            &      width (m \AA\ ) &dex& \\
\hline
  4607.327* &  Sr I &  0.000 & -0.57 &   43.9  &     2.02&10 \\
 5289.815* & Y II      &  1.033 & -1.85 &  70.4  &     1.52& 11\\
  5662.925 &       &  1.944  & 0.16 &  135.5  &     1.74& 12\\
  4934.076 &  Ba II &  0.000 & -0.15 &  414.6  &     1.38&5 \\
  6141.713* &    &  0.704&  -0.08 &  287.2  &     1.34& 5\\
  6496.897 &    &  0.604&  -0.38 &  298.0  &     1.47& 5\\
  4526.111 &  La II &  0.772 & -0.77 &   60.1  &     0.49& 13\\
 4921.776* &     &  0.244 & -0.68 &   94.6  &     0.09&14 \\
  4460.207 &  Ce II &  0.477&   0.17 &  146.3  &     1.49& 15\\
  4527.348 &    &  0.320 & -0.46&   144.3  &     1.79&15 \\
  4451.981 &  Nd II &  0.000 & -1.34&    38.3  &     0.10& 16\\
  4811.342 &     &  0.063 & -1.14 &   50.9  &     0.07&16 \\
  5212.361* &           &  0.204 & -0.87 &   61.5  &   0.04& 16\\
  5293.163 &          &  0.823 & -0.00 &   71.7  &     0.22&17 \\
 6645.064* & Eu II &   1.380 &  0.20 &  synth  & -0.40    & 18\\
   
  \hline
\end{tabular}

*Abundances are also  derived by spectral synthesis calculations of these lines. The results are presented in Table 5.\\
1.  \citet{kurucz75}, 2. \citet{laughlin74}, 3.  \citet{lincke71}, 4. Anderson et al. (1967), 5. \citet{miles69}, 6. \citet{martin88}, 7. \citet{bridges74}, 8. \citet{fuhr88}, 
 9.  \citet{kurucz88}, 10. \citet{corliss62},  11. \citet{hannaford82}, 12. \citet{cowley83}, 
13. \citet{andersen75}, 14. \citet{corliss62}, 15. \citet{mceachran71}, 
16. \citet{meggers75}  17. \citet{ward85}, 18. \citet{biemont82}\\
\end{table*}
}
\begin{table*}
\centering
{\bf  Table 4: Derived atmospheric parameters of CD$-$27~14351}\\
\begin{tabular}{lccccc}
\hline
Star Name.   & $T_{eff}$  &  log g  &  $\xi $ &[Fe I/H] &[Fe II/H]    \\
             &   K        &         &  km s$^{-1}$ &      &             \\
\hline
CD$-$27~1435&4335 &0.5 &2.42& $-$2.62 & $-$2.57\\
\hline
\end{tabular}
\end{table*}

\section{Results and discussions:  elemental abundances}
Elemental abundances are  derived from  the measured  equivalent 
widths of lines due to neutral and ionized elements if clean unblended lines are present.
We have also performed  spectral synthesis calculations to derive the 
elemental abundances for a few 
elements considering the hyperfine splitting effects. These elemental 
abundances along with the abundance ratios  with respect to iron are 
presented in table 
5. We have also calculated the [ls/Fe], [hs/Fe] and [hs/ls] values,
where ls represents light s-process elements Sr and Y and hs
represents heavy s-process elements Ba, La, Ce,  and Nd.
These estimates are presented in table 6.
  
\subsection{Carbon, Nitrogen, Oxygen}
Several strong molecular features of carbon are clearly noticeable  throughout 
the spectrum  of 
CD$-$27~14351. We have calculated the carbon abundance from the spectral 
synthesis 
of C$_2$ band at 5635  \AA\, (figure 3, middle panel). The 
 C$_2$ band at 5165  \AA\, is found to be saturated.   
Carbon  shows a large enhancement with [C/Fe] value of 
2.89 as estimated   from the C$_2$ molecular band at 5635  \AA\,. 
Abundance of carbon  derived from the CH band at 4300  \AA\, 
(figure 3,  lower panel) also shows similar value with [C/Fe] = 2.88. 
For the line lists of CH, C$_2$, CN  and $^{12}$C$^{13}$C we have consulted 
\citet{masseron14,brooke13,sneden14} and  
\citet{ram14}. Using the  estimated carbon abundance,  
nitrogen abundance was derived from spectral synthesis calculation 
of CN band with band head  at 4215 \AA\, (figure 4). 
Estimated  nitrogen  abundance  ${\sim}$  7.1 dex gives [C/N] = 1.0 
for this object. 
We could not detect 
any oxygen lines in the spectrum   due to severe distortion in the 
spectrum.  
We have  estimated  the $^{12}$C/$^{13}$C ratio from the $^{12}$CN 
and $^{13}$CN features near 8005  \AA\, (figure 3, top panel). 
Estimated   $^{12}$C/$^{13}$C = 10.1 for this object. Such a low
value is not unreasonable as the 
 estimated  log g value of  CD$-$27~14351 indicates that the object
 is  an evolved red giant
 star in which the CN cycled materials are expected to mix up into 
the atmosphere making the $^{12}$C/$^{13}$C ratio low.

\begin{figure}
\centering
\includegraphics[height=12cm,width=12cm]{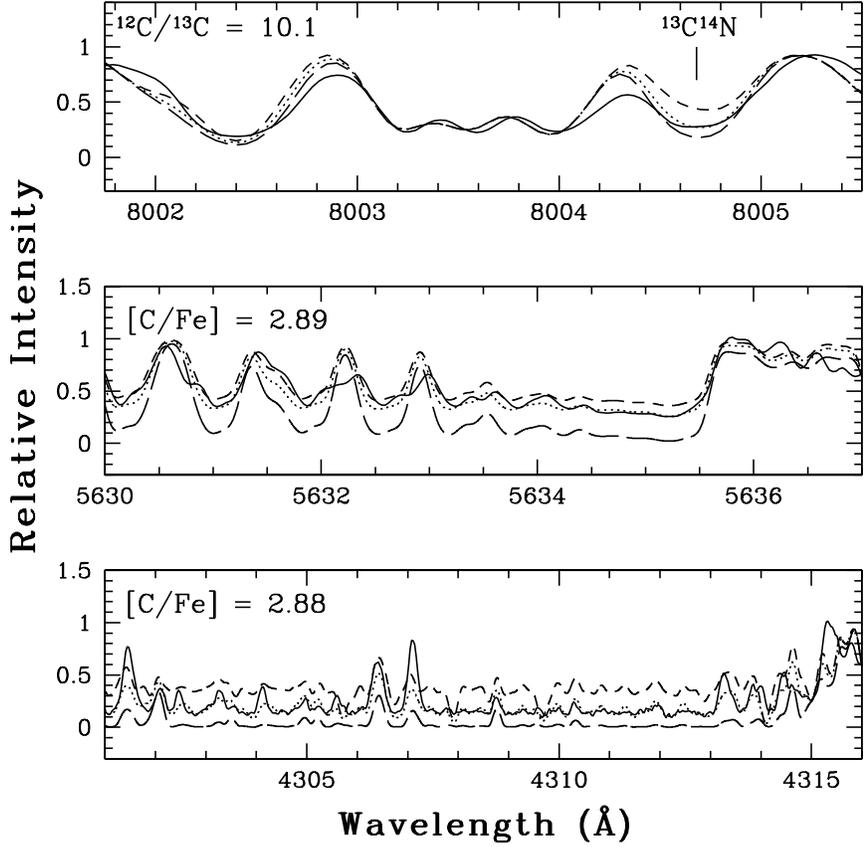}
\caption {Lower panel: The spectral synthesis fits of the CH feature around
 4315  \AA\,. 
Middle panel:  
The spectral synthesis of $C_2$ band around  5635  \AA\,. 
In both the panels the dotted lines indicate the synthesized spectra 
and the solid
lines indicate the observed line profiles. \ 
Two alternative synthetic spectra
for [C/Fe] = +0.3 (long-dashed line) and [C/Fe] = $-$0.3 (short-dashed
line) are shown to demonstrate the sensitivity of the line strength to the
abundances. 
Top panel: The spectral synthesis fits of the CN features around  8005  \AA\,  
obtained with  the adopted N abundance and  $^{12}$C/$^{13}$C ${\sim}$ 10.1 
(dotted curve). The observed spectrum is shown by  a solid curve. 
 Two alternative fits with  $^{12}$C/$^{13}$C ${\sim}$ 25 
(short-dashed line) and 5 (long-dashed line) are  shown to illustrate
the sensitivity of the line strengths to the isotopic carbon  abundance 
ratios. }
%label fig3
\end{figure}

\begin{figure}
\centering
\includegraphics[height=12cm,width=12cm]{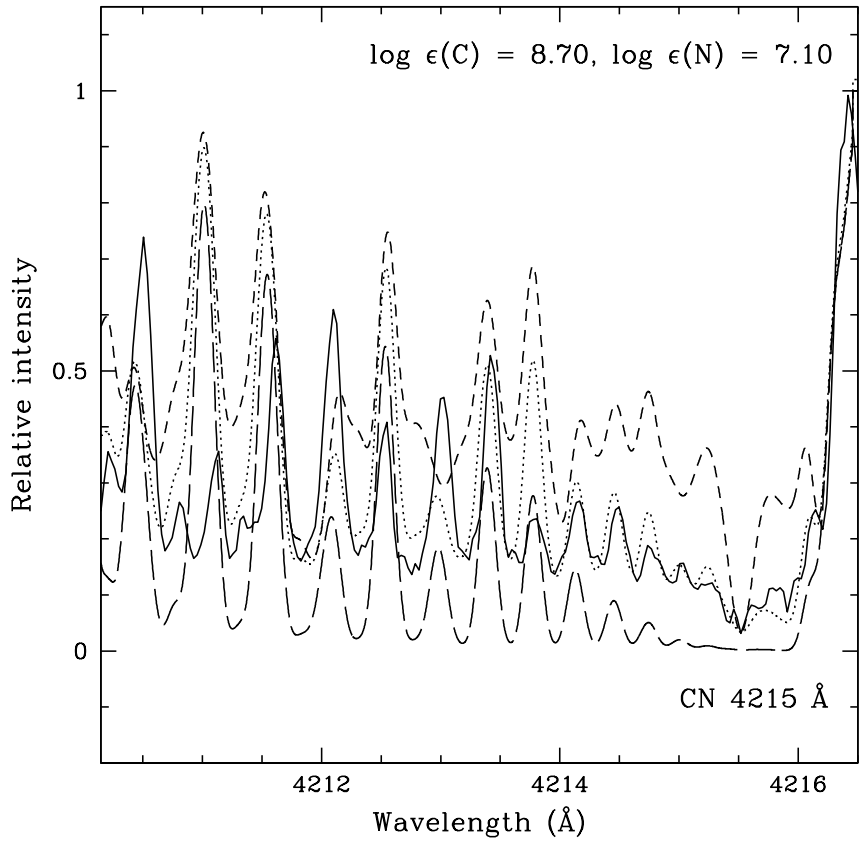}
\caption{ Spectral synthesis fits of CN band around  4215 \AA\,. The best
fit obtained  with  a carbon 
abundance of 8.7 dex and $^{12}$C/$^{13}C$ ${\sim}$ 10.1   returns a 
nitrogen abundance of 7.1 dex  (dotted lines). The solid line
corresponds to the observed spectrum. Two alternative 
plots with long-dash and short-dash are shown with [N/Fe] = $\pm$0.3 from 
the adopted value.  }
%label fig4
\end{figure}

{\footnotesize
\begin{table*}
\centering
{\bf Table 5: Elemental abundances in CD$-$27~14351}\\
\begin{tabular}{|l|c|c|c|c|c|}
\hline
 &       &                         &                &        &        \\
 &    Z  &   solar $log{\epsilon}^a$ & $log{\epsilon}$& [X/H]  & [X/Fe] \\
 &       &                         &                  &        &        \\
\hline
C {\sc i}  &   6      &   8.43  &  8.70$\pm$0.20  &   0.27  &  2.89 \\
N {\sc i}  &   7      &   7.83  &  7.10$\pm$0.20  &   -0.73  &  1.89 \\
Na {\sc i}  &  11     &   6.24   &  4.89$\pm$0.20  &   -1.35  &  1.27 \\
Mg {\sc i}  &  12     &   7.60   &  5.61$\pm$0.40  &   -1.99  &  0.63  \\
Ca {\sc i}  &  20     &   6.34   &  4.51$\pm$0.18  &   -1.83  &  0.79\\
Ti {\sc i}  &  22     &   4.95   &  3.25$\pm$0.18  &   -1.70  &  0.92\\
Ti {\sc ii} &  22     &   4.95   &  3.21$\pm$0.16  &   -1.74  &  0.83\\
Cr {\sc i}  &  24     &   5.64   &  2.84$\pm$0.17  &   -2.80  &  -0.18\\
Fe {\sc i}  &  26     &   7.50   &  4.87$\pm$0.15  &  -2.63& -\\
Fe {\sc ii} &  26     &   7.50   &  4.93$\pm$0.09  &  -2.57& -\\
Sr {\sc i}  &  38     &   2.87   &  2.02$\pm$0.20  &   -0.85  & 1.77\\
Sr {\sc i}$^*$  &  38     &   2.87   &  1.98$\pm$0.20  &   -0.89  & 1.73\\
Y {\sc ii}  &  39     &   2.21   &  1.63$\pm$0.12  &   -0.58  &  1.99\\
Y {\sc ii}$^*$  &  39     &   2.21   &  1.55$\pm$0.20  &   -0.66  &  1.91\\
Ba {\sc ii} &  56     &   2.18   &  1.40$\pm$0.07  &  -0.78   & 1.79\\
Ba {\sc ii}$^*$ &  56     &   2.18   &  1.38$\pm$0.20  &  -0.8    & 1.77\\
La {\sc ii}   &  57     &   1.10 &  0.29$\pm$0.18  &-0.81 &  1.76\\
La {\sc ii}$^*$ &  57     &   1.10   &  0.10$\pm$0.20  &   -1.00  & 1.57\\
Ce {\sc ii} &  58     &   1.58   &  1.43$\pm$0.30  &   -0.15  & 2.63\\
Nd {\sc ii} &  60     &   1.42   &  0.11$\pm$0.08  &   -1.31  & 1.26\\
Nd {\sc ii}$^*$ &  60     &   1.42   &  $-$0.05$\pm$0.20  &   -1.47  & 1.10\\
Eu {\sc ii}$^*$ &  63     &   0.52   &  $-$0.40$\pm$0.20  &   -0.92   & 1.65\\
\hline 
\end{tabular}

$^{a}$ \citet{asplund09} \\
$^*$ Abundance estimates are from  spectral synthesis calculation. \\
\end{table*}
}
\subsection{Sodium (Na) and Aluminum (Al)}
We have derived the Na abundance from Na I line at 5688.22 \AA\ . Estimated
 Na abundance is high  with 
[Na/Fe] = 1.27. Since this  line  is free from NLTE effects 
\citep{baumueller97}
we have not applied NLTE correction to the derived abundance.  A few 
very metal-poor stars are  also known to show such high abundances of Na 
\citep{aoki02a,aoki02b,aoki07,aoki08}.  Abundance of Al could  not  be estimated
 in CD$-$27~14351.
\subsection{Abundances of ${\alpha}$- elements}
 We have estimated  [Mg/Fe] ${\sim}$ 
 0.64 from three  Mg I lines at 4571.1,   5172.68 and 5528.41 \AA\ . 
This value is slightly higher than  the general trend 
of [$\alpha$/Fe] = 0.4 noticed in metal-poor stars. 
We have measured three clean Ca I  lines in the spectrum of 
CD$-$27~14351. Similar to Mg, calcium also shows large enhancement with 
[Ca/Fe] value 0.79.   
Ti shows  enhancement with  [Ti I/Fe] and [Ti II/Fe] values 0.92 and 
0.83 respectively.  \citet{johnson02}  have also found that Ti I and Ti II
are discrepant in metal-poor stars, likely because of NLTE effects.
  These estimates show that the object is a very metal-poor star with high
  abundances of  $\alpha$ elements. 
\subsection{Fe-peak elements}
Two Cr I lines at 5247.57 and 5348.31 \AA\, are used to derive the Cr
abundance. Cr shows under abundance with [Cr/Fe] value $-$0.18. 
We have detected  many Ni lines in the spectrum of CD$-$27~14351. 
But none of them are usable for  abundance calculation due to 
distortion and severe  line blending. 
\subsection{Abundances of  neutron-capture elements}
Abundances of  neutron-capture elements are derived 
by the equivalent width measurements as well as by the spectral synthesis 
calculations. We have accounted for any possible  blending due to other
elements  in the linelists  of the respective lines used for spectral
synthesis calculation.  
The derived abundances along with the error estimates  are listed in table 5.   
Abundance of Sr is derived from the Sr I line at 4607.7 \AA\,. Sr is
found to be  enhanced with [Sr/Fe] ${\sim}$ 1.75.  Y also shows 
an enhancement with [Y/Fe] ${\sim}$ 1.95. The spectrum synthesis fits
are shown in figure 5.
We could not measure any Zr lines in the spectrum of CD$-$27~14351.
Ba abundance calculated by measuring the equivalent widths of Ba II lines 
at 4934,  6141 and 6496  \AA\ give a [Ba/Fe] 
value of 1.79. The spectrum synthesis of Ba II line at 6141.713 \AA\ (figure 6)
 considering the hyperfine splitting contributions from \citet{mcwilliam98} gives [Ba/Fe] ${\sim}$ 1.77. 
 We have derived La abundance from spectrum
synthesis calculation of La II line at 4921.77  \AA\ considering hyperfine 
components from \citet{jonsell06} which gives [La/Fe] ${\sim}$ 1.57. 
Cerium abundance derived from two Ce II lines at 4460.2 and 4527.3 \AA\  
indicates  
a large  enhancement of Ce with [Ce/Fe] ${\sim}$ 2.63. Nd abundance 
calculated by the 
equivalent width measurement of four Nd II lines gives [Nd/Fe] ${\sim}$ 1.26. 
We have also derived the Nd abundance  by  spectral synthesis calculation
 of Nd II line at 5212.3 \AA\ (figure 5); this gives a value of 1.1 for  
[Nd/Fe]. 
We have derived  Eu abundance from Eu II line at 6645.130
\AA\ by considering the hyperfine components from \citet{worley13}. 
Although the right wing of this Eu line is found to be blended with 
Fe I line at 6645.36 \AA\, and  a Tb I line at 6645.37 \AA\, we could get a 
proper fit for the left wing with the adopted Eu abundance of $-$0.40 dex 
(figure 6). 
The  adopted  Eu abundance gives [Eu/Fe] ${\sim}$ 1.65. We could not detect 
 Eu lines at 4129 and 6437 \AA\,.

\begin{figure}
\centering
\includegraphics[height=12cm,width=12cm]{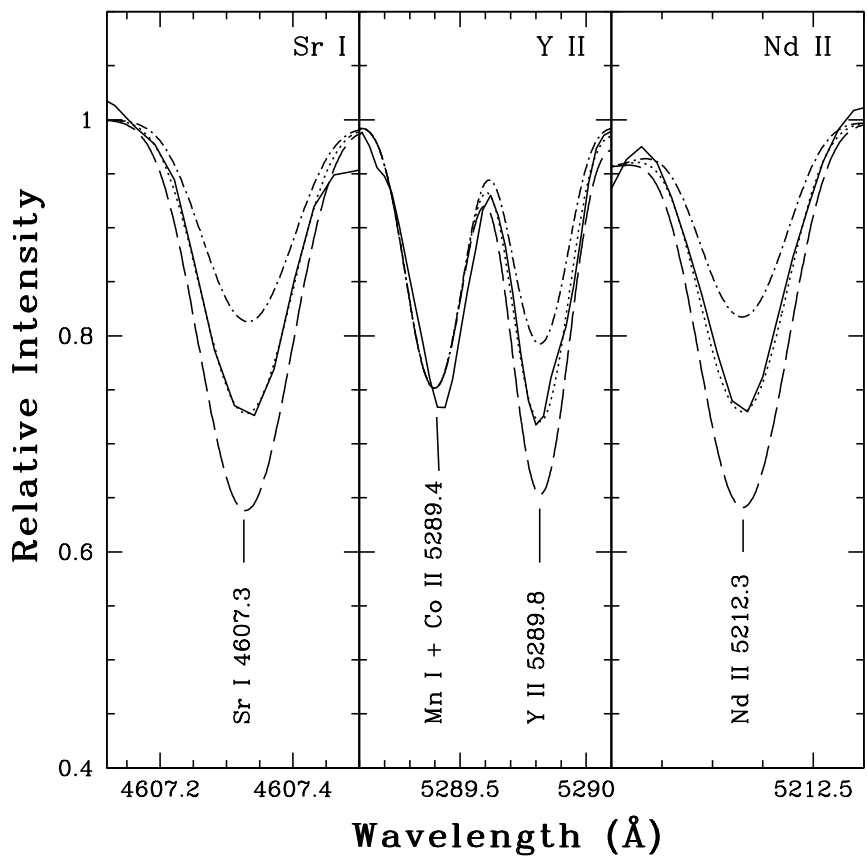}
\caption{Spectral-synthesis fits of Sr I line at 4607.3 {\rm \AA}, Y II line at 5289.8 {\rm \AA} and Nd II 
line at 5212.3 {\rm \AA} are shown. 
The dotted lines indicate the synthesized spectra and the solid
lines indicate the observed line profiles. Two alternative synthetic spectra
for [X/Fe] = +0.3 (long-dashed line) and [X/Fe] = $-$0.3 (short-dashed
line) from the adopted  value are shown to demonstrate the sensitivity 
of the line strength to the abundances.}
%label fig5
\end{figure} 

 \begin{figure}
\centering
\includegraphics[height=12cm,width=12cm]{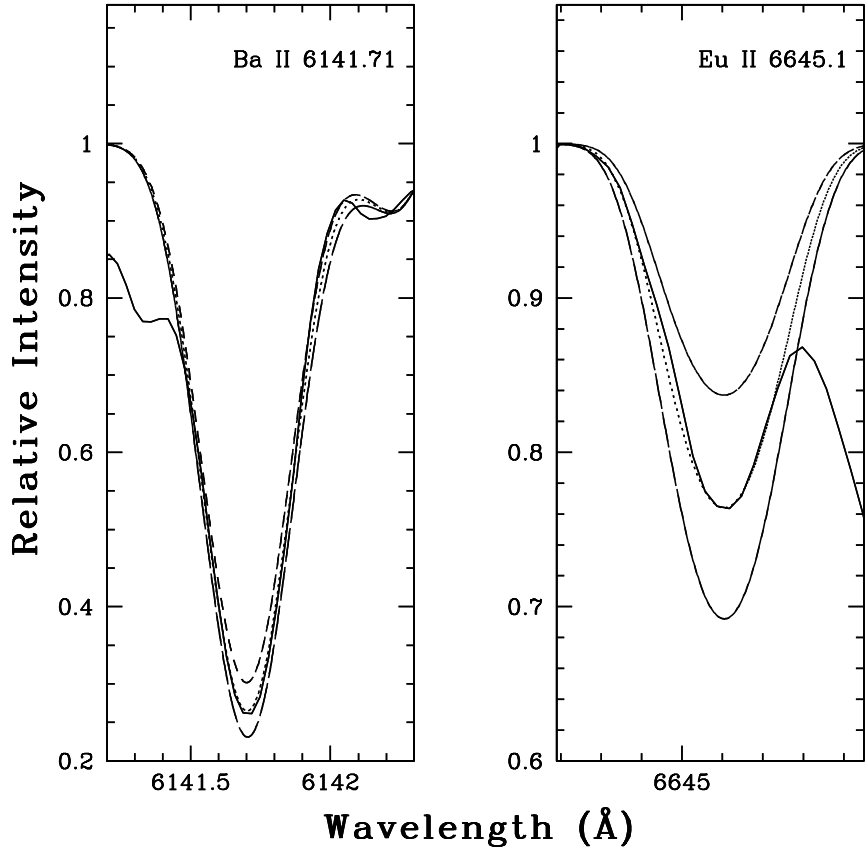}
\caption{Spectral-synthesis fits of Ba II line at 6141.71 {\rm \AA} and 
Eu II line at 6645 {\rm \AA}. In both the panels 
the dotted lines indicate the synthesized spectra and the solid
lines indicate the observed line profiles. Two alternative synthetic spectra
for [X/Fe] = +0.3 (long-dashed line) and [X/Fe] = $-$0.3 (short-dashed
line) from the adopted  value are shown to demonstrate the sensitivity 
of the line strength to the abundances.}
%label fig6
\end{figure}

 \begin{table*}
\centering
{\bf Table 6: Abundance ratios of light  and heavy s-process elements}\\
\begin{tabular}{lcccc}
\hline
Star Name & [Fe/H]  & [ls/Fe] & [hs/Fe] & [hs/ls]   \\
CD$-$27 14351& $-$2.62& 1.82& 1.77&  $-$0.05\\ 
\hline
\end{tabular}

[ls/Fe] and [hs/Fe] are calculated using the abundance values obtained from
spectral synthesis calculations as given in table 5.
\end{table*}
  \begin{table*}
\centering
{\bf Table 7: Abundance Uncertainties}\\
\begin{tabular}{cccccc}
\hline
Star & Standard  & $\delta T_{eff}$&$ \delta$ log g &$ \delta \xi$ & Total\\
     &    error            & $\pm$100K         & $\pm $0.1    & $\pm$0.3 km/s &  Uncertainty \\  
\hline

CD$-$27~14351&0.15&  0.08& 0.01& 0.09&0.17\\
\hline
\end{tabular}
\end{table*}
\section{Uncertainty in elemental abundance}
We have derived the uncertinities in elemental abundance estimates by varying
 the stellar atmospheric  parameters $T_{eff}$, log g 
and microturbulence in the model atmosphere. We have calculated the 
uncertainty due to  temperature by varying the 
temperatures by $\pm$ 100 K and recalculated the Fe abundance. Similarly by
 varying the log g value by $\pm$ 0.1 and micro turbulent velocity by 
$\pm$ 0.1 km s$^{-1}$, we have calculated the corresponding uncertainties in 
abundances due to these changes. 
These values, along with the standard error are listed in table 7. The 
total uncertainty is calculated using the standard equation of error 
calculations     
           \begin{equation}
E_r = \sqrt{{E_{r1}}^2 + {E_{r2}}^2 + {E_{r3}}^2 + {E_{r4}}^2}
\end{equation}
We have assumed this value as the minimum error in the derived abundances.

\section{Conclusions}
A detailed chemical composition study of CD$-$27 14351 based on a high 
resolution FEROS spectrum revealed many important features of this object.
The object is characterized by a large enhancement in carbon; a similar
enhancement in nitrogen is also evident from our analysis. Such high 
abundances of carbon and nitrogen are also seen in other CEMP stars
(figure 7);  these two elements however show a large scatter with 
respect to metallicity. The ${\alpha}$-elements Mg, Ca, Ti and Fe-peak
element Cr show similar abundances as are generally noticed in CEMP stars;
 the high abundance of Na is also a common feature of many
CEMP stars (figure 8). A comparison of our estimated abundance ratios 
of neutron-capture elements with those  of \citet{allen12,goswami06,goswami10,goswami16} and \citet{masseron10},
for CEMP stars is shown in figure 9. A similar comparison of the 
neutron-capture elements abundance ratios observed in CD$-$27 14351
with their counterparts observed in CEMP-s and CEMP-r/s stars alone
from \citet{jonsell06} clearly shows that Sr, Y and Ce are highly enhanced 
whereas Ba, La and Eu are similarly enhanced as seen in the case of 
 CEMP-r/s stars.
Large enhancement of s-process elements along with the enhancement in Mg
indicates the operation of $^{22}$Ne($\alpha$,n)$^{25}$Mg as the main
neutron source in CEMP-r/s stars \citep{masseron10}. As suggested
by \citet{gallino98} and \citet{goriely00}  neutron
density associated with this reaction  favours production of the 
s-process element Ce and also the r-process element  Eu. 
 Highly enhanced abundances of these two elements observed in
this star support  this idea. However, \citet{goriely05} suggest 
that a $-$ve
value of [La/Ce] indicates the operation of $^{13}$C($\alpha$,n)$^{16}$O 
reaction, and,  a value of [La/Ce] within  the  range  0.2 to 0.4 
 indicates  the operation of
$^{22}$Ne($\alpha$,n)$^{25}$Mg reaction.
Our estimated [La/Ce] = $-$1.06 would   therefore imply the reaction 
  $^{13}$C($\alpha$,n)$^{16}$O  to be the main source of neutrons.  
Estimated [Ba/Eu] (= 0.12) indicates that the object
can be placed in CEMP-r/s group.   
Although the binarity among CEMP-s stars are well known,  
there are a very few studies (e.g. \citet {abate16} ) 
focussed on the orbital properties of CEMP-r/s stars. Inspite of 
several efforts in literature the origin of CEMP-r/s stars still 
remains far from clearly understood.  However in future, it would
be worthwhile to investigate if the observed abundance pattern 
in CD~$-$27 14351 could arise from nucleosynthesis of i-process.
 Our observational results are
expected to help constrain  theoretical modelling  providing
insight and better understanding of the origin of such objects.
 
\begin{figure}
\centering
\includegraphics[height=12cm,width=12cm]{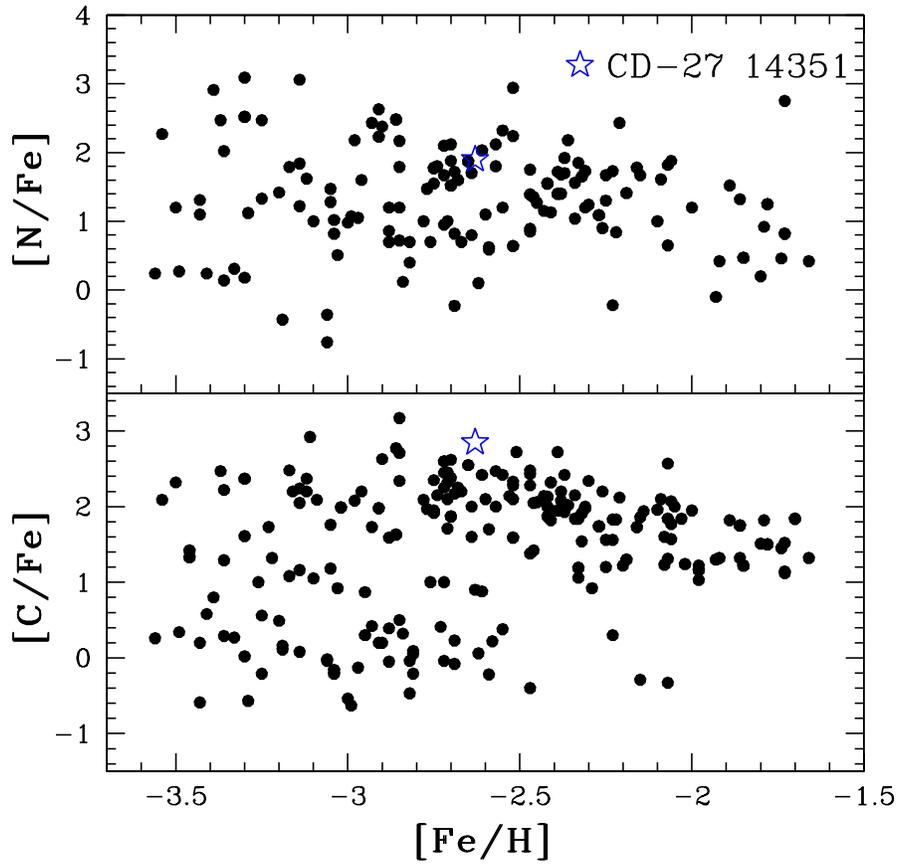}
\caption {This figure shows a comparison of the  [C/Fe]  
(lower panel) and [N/Fe] (upper panel) values   observed in CD$-$27~14351 
 (shown with a star symbol in both the panels)
with their counterparts (solid circles)  in CEMP stars from  literature 
(e.g., \citet{masseron10}). }
%label fig7
\end{figure}

 \begin{figure}
\centering
\includegraphics[height=12cm,width=12cm]{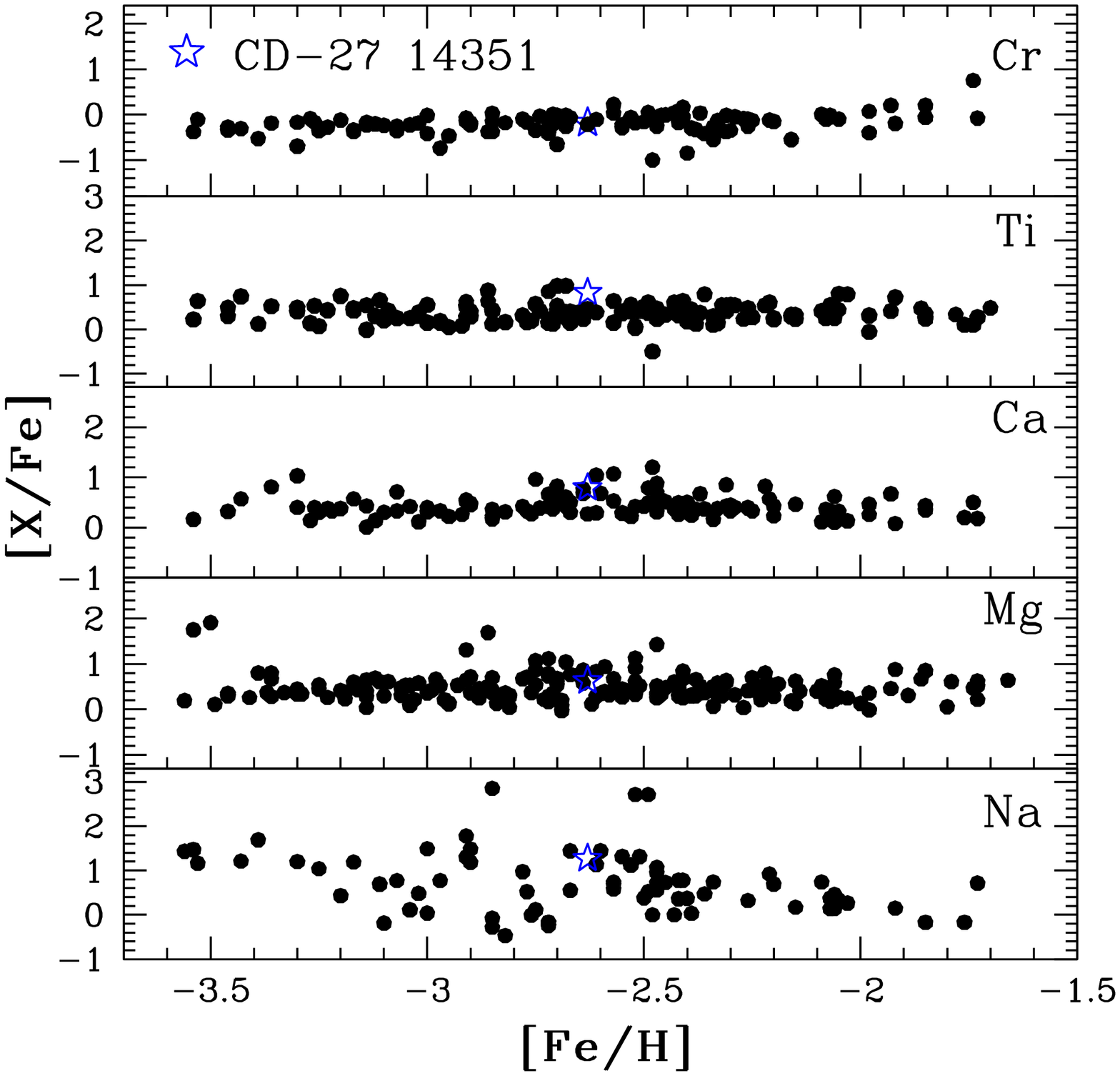}
\caption{Comparison of the observed light element abundances of CD$-$27~14351 
(indicated by a star symbol) with their counterparts  in CEMP stars 
available in  literature (i.e., \citet {allen12,goswami06,goswami10,goswami16,masseron10} ). }
%label fig8
\end{figure}

\begin{figure}
\centering
\includegraphics[height=12cm,width=12cm]{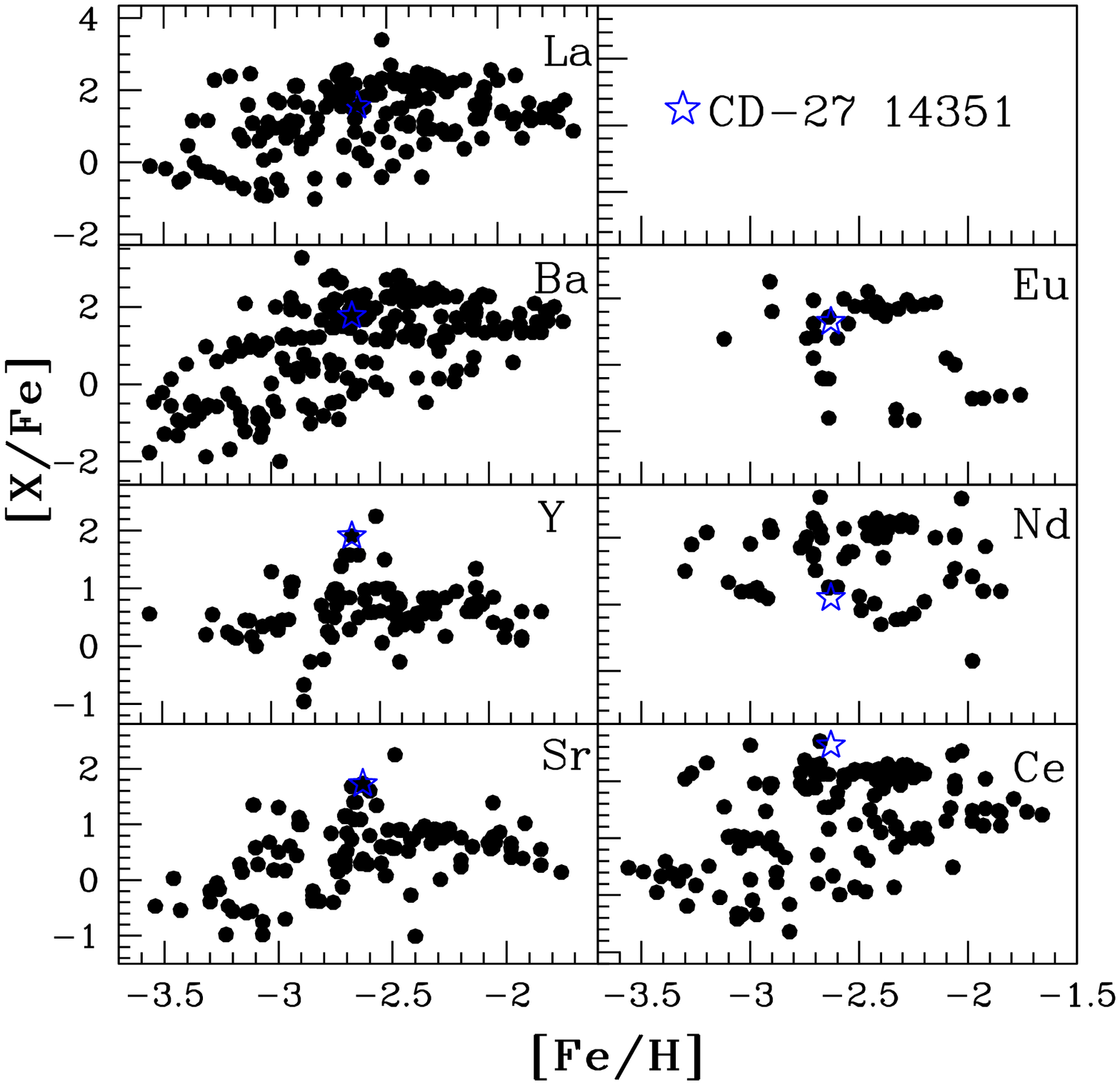}
\caption{ Comparison of the observed heavy element abundances in
 CD$-$27~14351 (indicated by a star symbol) with their counterparts  
in CEMP stars available in literature (i.e., \citet{allen12,goswami06,goswami10,goswami16,masseron10}). }
%label fig9
\end{figure}

\begin{figure}
\centering
\includegraphics[height=12cm,width=12cm]{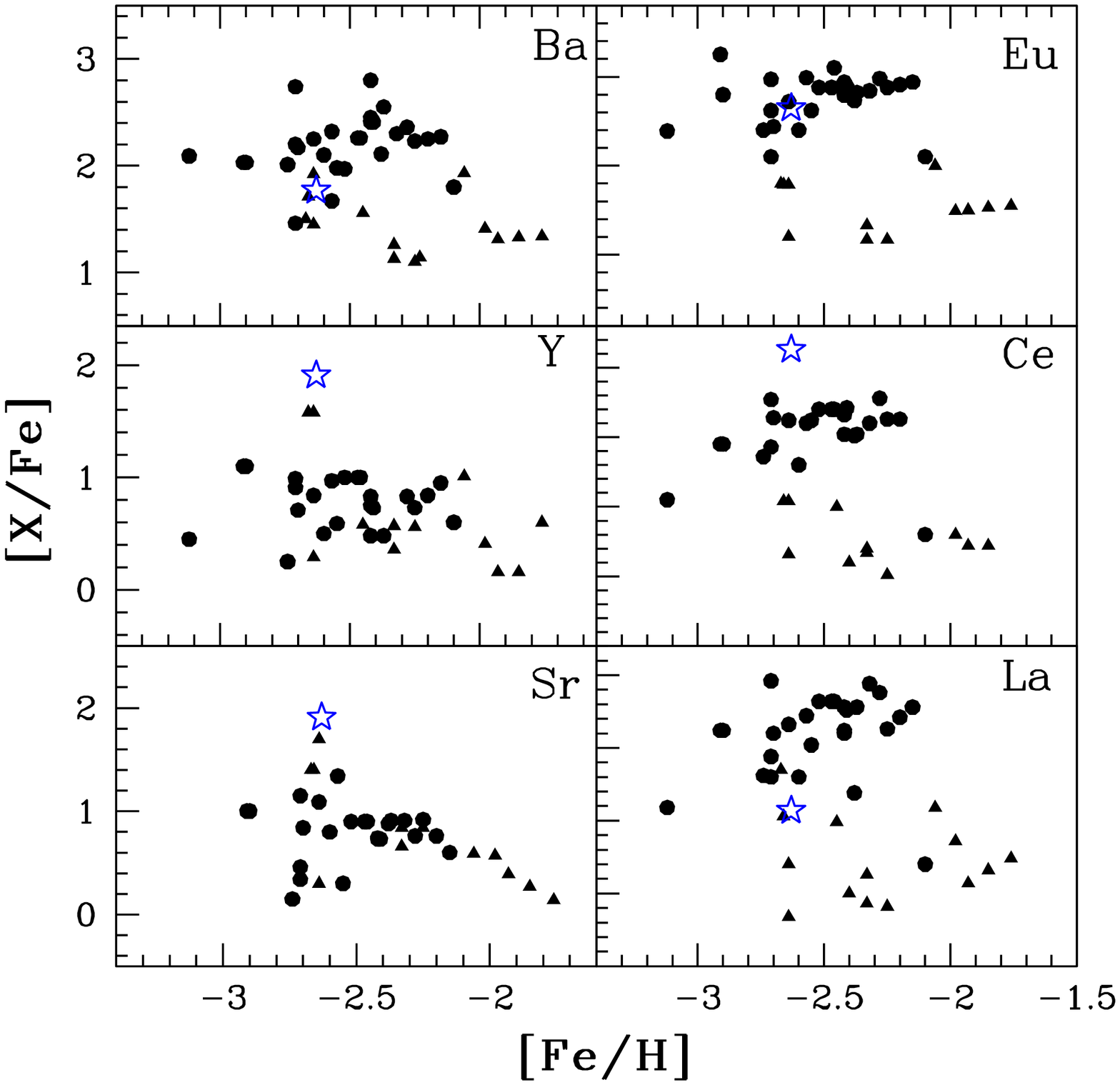}
\caption{comparison of the observed heavy element abundances in 
CD$-$27~14351 (indicated by a star symbol) with  values from 
other CEMP-r/s (solid circle) and 
CEMP-s stars (solid triangle) from \citet{jonsell06}.}
%label fig10
\end{figure}

\vskip 0.3cm
\noindent
{\it Acknowledgements}\\
AG, the PI of the HCT observing proposal and DK  would like
to  thank  the staff at IAO and at the remote control station at CREST,
Hosakote, for assistance during the observations.
This work made use of the SIMBAD astronomical data base, operated at CDS,
Strasbourg, France, and the NASA ADS, USA.  Funding from the DST project
SB/S2/HEP-010/2013 is gratefully acknowledged. This work was partly
supported by the European Union FP7 programme through ERC grant number 
320360.
%\begin {thebibliography}

\bibliographystyle{aa}
\bibliography{reference}

%\end {thebibliography}
\end{document}